\documentclass{ptephy}
\usepackage{color}

\begin{document}

\title{Jet-induced jammed states of granular jet impacts}

\author{\name{Tomohiko G. Sano}{1\ast} and  \name{Hisao Hayakawa}{2\dagger}}

\address{\affil{1,2}{Yukawa Institute for Theoretical Physics, Kyoto University Kitashirakawa Oiwakecho, Sakyo-ku, Kyoto 606-8502 Japan}
\email{tomohiko@yukawa.kyoto-u.ac.jp, $^\dagger$E-mail: hisao@yukawa.kyoto-u.ac.jp}}

\begin{abstract}%
The impacts of granular jets for both frictional and frictionless grains in two dimensions are numerically investigated. 
A dense flow with a dead zone emerges during the impact. 
From our two-dimensional simulation, we evaluate the equations of state and the constitutive equations of the flow. 
The asymptotic divergences of pressure and shear stress similar to the situation near the jamming transition appear for the frictionless case, while their exponents are smaller than those of the sheared granular systems, and are close to the extrapolation from the kinetic theoretical regime. 
In a similar manner to the jamming for frictional grains, the critical density decreases as the friction constant of grains increases.
For bi-disperse systems, the effective friction constant defined as the ratio of shear stress to normal stress, monotonically increases from near zero, as the strain rate increases. 
On the other hand, the effective friction constant has two metastable branches for mono-disperse systems because of the coexistence of a crystallized state and a liquid state.
\end{abstract}

\subjectindex{J44, J01, A56}

\maketitle

\section{Introduction}
Non-equilibrium phenomena induced by impacts have been extensively studied in various contexts, such as nuclear reactions \cite{nuclei1,nuclei2,nuclei3}, nanotechnology \cite{nc1,nc2}, water-bells \cite{wbell1,wbell2} and granular flows \cite{gjet,crater1,crater2,crater3,sano_hayakawa1,sano_hayakawa2,guttenberg,wendy,huang, wendy2, wendy3}. Crater morphology is studied via an impact process of a free-falling water drop or a grain onto a granular layer \cite{crater1,crater3}, while a sinking grain produces a sand jet \cite{crater2}. The impact of a granular jet on a target produces a sheet-like scattered pattern or a cone-like pattern, depending on the ratio of the target diameter and the jet diameter \cite{gjet}, which is also found in water-bell experiments with low surface tensions \cite{wbell1,wbell2}.

Cheng {\itshape et al.} suggested that the fluid state of a granular jet after an impact is similar to the Quark Gluon Plasma(QGP), which behaves as a perfect fluid through their experiment \cite{gjet}. 
Recently, we reported that the shear viscosity during the impact is well described by the kinetic theory of the granular gas \cite{garzo, torquato, jenkins, yoon,saitoh, jenkins_richman}, though the small shear stress observed in the experiment is reproduced through our three-dimensional (3D) simulation \cite{sano_hayakawa1, sano_hayakawa2}.
 Because the shear viscosity, at least, for 3D is not anomalous, the correspondence between a granular flow and QGP would be superficial. 

To discuss the fluid state of granular jets, we need to know the details of rheology of moderate dense granular flows.
A typical situation of the study for a dense granular flow is the flow on an inclined plane \cite{bagnold1,bagnold2,bagnold3}. Bagnold proposed the constitutive equation for dense granular flows that the shear stress is proportional to the square of the shear rate \cite{bagnold1}, so called Bagnold's scaling, which has been verified experimentally \cite{bagnold2} and numerically \cite{bagnold3,incl1} under several conditions such as the flow down an inclined plane. 
Dense granular flows, however, have more variety of rheological constitutive equations for flows on inclined planes \cite{jop,poliq,poliq_eq,GDR,hill}. Conventional one would be the constitutive equation presented by Jop and coworkers \cite{jop}, where the effective friction constant, defined as the ratio of shear stress to pressure, saturates from a static value at zero shear rate to a maximum value as the shear rate increases.
The power-law friction law, which is also different from Bagnold's scaling for dense granular flows, is proposed via extensive simulations \cite{cruz,hatano,hatano_jps,ohern}.

A granular system has rigidity above a critical value of density $\phi_{\rm J}$ and does not have any rigidity below $\phi_{\rm J}$. 
This sudden change of the rigidity is known as the jamming transition \cite{liu_nagel, OH1, Teitel, OHL, hatano_jam,OH2, OH3, garcia, losert,OH4,Teitel0,tighe,nord,hatano_jam1}. 
The jamming is not only investigated in systems of grains, but also that of colloidal suspensions \cite{j_colloid} or foams \cite{j_foam} . 
Here, $\phi_{\rm J}$ decreases as the friction constant $\mu_{\rm p}$ of grains increases.
Moreover, it seems that there are two fictitious jamming points in addition to the true jamming point for finite $\mu_{\rm p}$ \cite{OH1}. Critical exponents of the divergence of the pressure and the shear viscosity near the transition are extensively discussed \cite{OH1, Teitel, OHL, hatano_jam,OH2, OH3, OH4,garcia, losert,Teitel0,tighe,nord,hatano_jam1}.

The aim of this paper is to investigate the rheological properties for two-dimensional (2D) granular jet impacts. Although some previous numerical studies on granular jets used 2D simulations to reproduce 3D experiments for the computational efficiency \cite{huang, guttenberg, wendy}, it is unclear whether the rheological properties in 2D granular jets are qualitatively the same as those in 3D. Therefore, to clarify the qualitative difference between 2D and 3D granular jets is necessary. Because grains are easily packed through the impact in 2D, the system would be near the jammed state. Thus, we can investigate rheological properties of very dense granular fluids after the impact of granular jet flow, which cannot be achieved by 3D simulations and experiments. As a result, correlated flows appear in 2D granular jets, while uncorrelated flows characterized by the granular kinetic theory is realized in 3D jets. There are another advantage for the visualization to use 2D system even for experiments to know detailed properties of particles in granular jets, such as contact networks (force chains) and the effect of crystallization for mono-disperse case. We also stress that it is easy to perform 2D or one layer experiments for granular jets. 

In this paper, we perform 2D simulations for the granular jet in terms of the discrete element method (DEM) \cite{DEM}. 
This paper is complementary to the previous 2D DEM study \cite{huang}, and hard core simulations supplemented by the simulation of a perfect fluid model \cite{wendy}. Indeed, although Huang {\it et al.} reported that the relevant role of the contact stress in a 2D granular jet, they were not interested in the critical behavior of jammed grains induced by the jet. Guttenberg suggested that the friction constant does not play a significant role, at least, in the scattering angle \cite{guttenberg}, while the effects on the jammed state induced by jets have not been studied in his paper.

This paper is organized as follows: After the introduction of our numerical model in Sec. \ref{model}, we analyze the profile of the local stress tensor, the area fraction and the granular temperature. We also discuss the rheology of the granular jets for the frictionless case in 2D to compare their behavior with the jamming transition for a bi-disperse frictionless case. The effect of the friction constant is discussed in Sec. \ref{section_friction}. In Sec. \ref{discussion}, our numerical results for a mono-disperse case are shown and the paper is concluded in Sec. \ref{concl}. In the Appendix A, we comment on the artificial burst-like flow in 2D, which appears in the case of large $\mu_{\rm p}$ for soft grains. In the Appendix B, we discuss the effect of the inhomogeneity of the temperature to Balgnold's scaling in terms of the method of Green's function.

\begin{figure}[h]
\centering
\includegraphics[scale = 0.5]{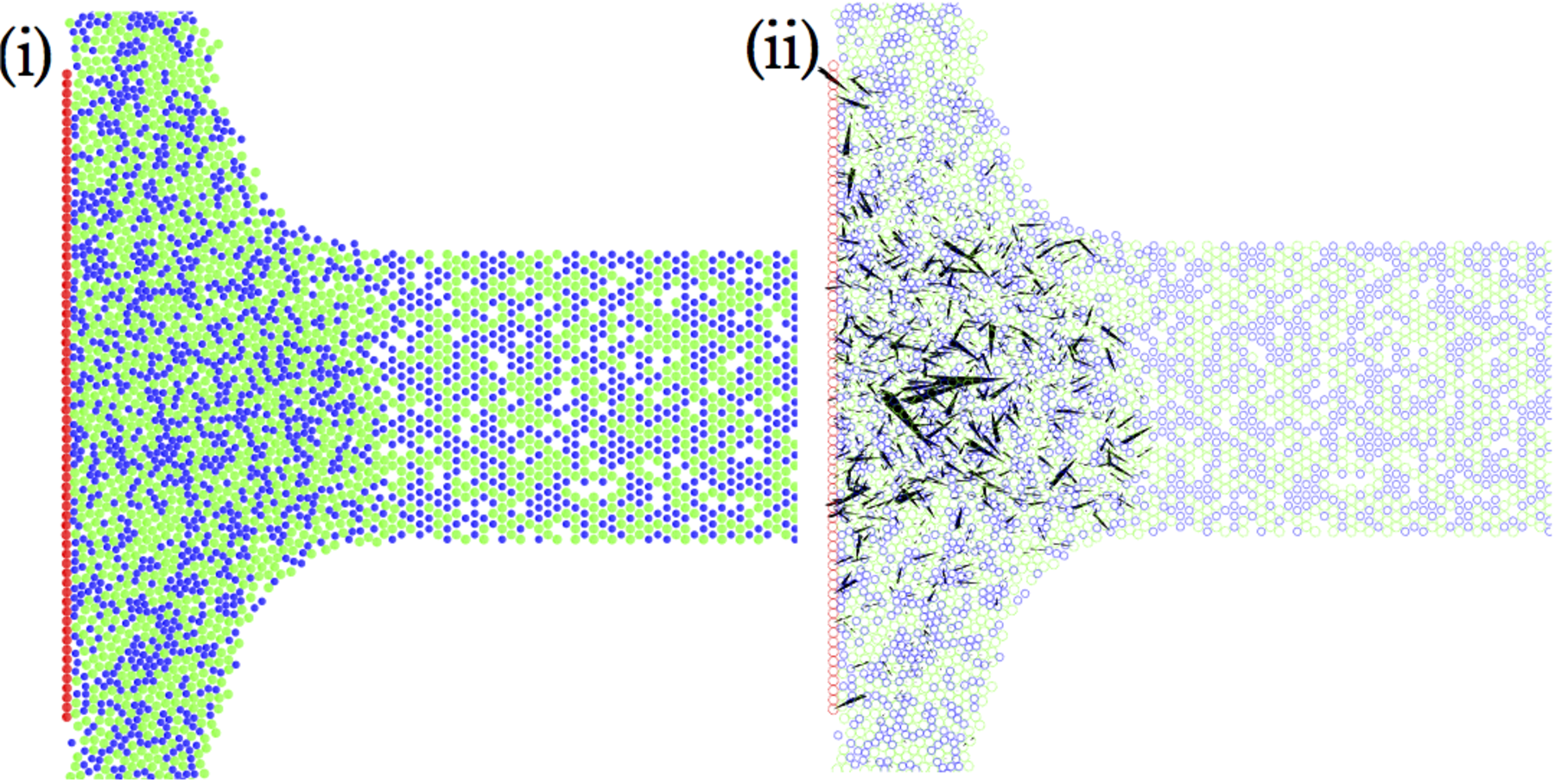}
\caption{(i) A typical snapshot of the simulation for the frictionless case with $\tilde{\phi}_0 = 0.90$. Blue particles and green particles denote grains with diameter $0.8d$ and $d$, respectively and red particles are wall-particles. (ii) The corresponding contact forces among grains are visualized as black colored arrows. The average coordination number $Z  \simeq 0.526$ and $71.5\%$ of particles are not in contact in the region $0<x\leq 10$ and $|y| < R_{\rm tar}$.}
\label{snap}
\end{figure}

\section{Model}\label{model}
We adopt DEM to simulate the jet \cite{DEM}. We mainly focus on bi-dispersed soft core particles of the diameter $d$ and $0.8d$ with the same mass $m$ to avoid the crystallization. 
When the particle $i$ at the position ${\bf r}_i$ and the particle $j$ at ${\bf r}_j$ are in contact, 
the normal force $F^{\rm n}  _{ij}$ is given by $F^{\rm n} _{ij}  \equiv  F_{ ij} ^{\rm (el)} + F_{ ij} ^{\rm (vis)}$ with  $F_{ ij} ^{\rm (el)}  \equiv  k_{\rm n}(R_i + R_j - r_{ ij})$ and $F_{ ij} ^{\rm (vis)}  \equiv -\eta_{\rm n}({\bf g}_{ ij} \cdot \hat{\bf r}_{ ij})$,
 where  $r_{ij}  \equiv  |{\bf r}_{i} - {\bf r}_{j}|$ and  ${\bf g}_{ij}  \equiv  {\bf v}_{ i} - {\bf v}_{ j}$ 
with the velocity ${\bf v}_i$ and the radius $R_i$ of the particle $i$.
 The tangential force is given by $F^{\rm t} _{ij} \equiv \min\{ |\tilde{F^{\rm t} _{ij}}|, \mu_{\rm p} F_{ij}^{\rm n} \}{\rm sgn}(\tilde{F}_{ij}^{\rm t}) $, 
where the sign function is defined to be ${\rm sgn}(x)=1$ for $x\ge 0$ and ${\rm sgn}(x)=-1$ for otherwise, 
$\tilde{F^{\rm t} _{ij}} \equiv k_{\rm t} \delta^t _{ij} - \eta_{\rm t} \dot{\delta}_{ij} ^{\rm t}$
with the tangential overlap $\delta^{\rm t} _{ij}$ and the tangential component of relative velocity $\dot{\delta}^{\rm t} _{ij}$ between $i$ th and $j$ th particles. We examine the value of $\mu_{\rm p}$ from $\mu_{\rm p} = 0.2$ to $1.0$. Here, we adopt parameters $k_{\rm n} = 4.98 \times 10^2 mu_0 ^2 /d^2$, $\eta_{\rm n} = 2.88 u_0 /d$, with the incident velocity $u_0$ for the frictionless case and $\mu_{\rm p} = 0.2$. The value $\mu_{\rm p} = 0.2$ is close to the experimental value for nylon spheres \cite{Rosato}. 
We use $k_{\rm n} =  1.99 \times 10^3 mu_0 ^2 /d^2, \eta_{\rm n} = 5.75 u_0 /d$  for $\mu_{\rm p} = 0.4$, and $k_{\rm n} =  7.96\times 10^2 mu_0 ^2 /d^2, \eta_{\rm n} = 10.15 u_0 /d$ for $\mu_{\rm p} = 1.0$. These sets of parameters imply that the duration times are, respectively, $t_{\rm c} = 0.10 d/u_0$ for the frictionless case and $\mu_{\rm p} = 0.2$, $t_c = 0.05 d/u_0$ for $\mu_{\rm p} = 0.4$ and $t_c = 0.01d/u_0$ for $\mu_{\rm p} = 1.0$, the restitution coefficient for a normal impact is unchanged ${e} = 0.75$ for the frictionless case and for all $\mu_{\rm p}$. The reason why we adopt these parameters for large $\mu_{\rm p}$ is that many overlaps among grains lead to the artificial burst-like flow, if we adopt the identical $t_{\rm c}$ to frictionless case, as is shown in the Appendix A. For the tangential parameters, we choose $k_{\rm t} = 0.2 k_{\rm n}, \eta_{\rm t} = 0.5 \eta_{\rm n}$.
We adopt the second-order Adams-Bashforth method for the time integration of Newton's equation with the time interval $\Delta t = 0.02 t_{\rm c}$. 

An initial configuration is generated as follows: We prepare a triangular lattice with distance between grains $1.1d$ and remove particles randomly to reach the desired density. 
We control the initial area fraction $\phi_0 / \phi_{\rm ini} \equiv \tilde{\phi_0}$ before the impact as $0.30 \leq \tilde{\phi_0} \leq 0.90$ with the initial area fraction before the removal $\phi_{\rm ini} = 0.612, 0.780$ for the bi-disperse and the mono-disperse case, respectively, and 8,000 particles are used. We average numerical data over the time $180.0 \leq tu_0/d < 300.0$ after the impact. 
The initial granular temperature, which represents the fluctuation of particle's motion, is zero. 
The wall consists of particles in one layer with the same diameter $d$ and the same mass $m$, which are connected to each other and with their own initial positions via the spring and the dashpot with the spring constant $k_{\rm w} = 10.0 mu_0 ^2 /d^2$ and the dashpot constant  $\eta_{\rm w} = 5.0 \eta_{\rm n}$, respectively. 

A typical snapshot of our simulation and that of the contact force network are shown in Fig. \ref{snap} (i) and (ii), respectively. Blue, green and red particles denote grains with diameter $0.8d$ and $d$, and wall-particles, respectively in Fig. \ref{snap} (i) and all of the corresponding contact force network among grains are visualized as black colored arrows in Fig. \ref{snap} (ii). It is easily found that the contact force network emerges during the impact. It should be noted that the average coordination number $Z \equiv \sum_{i \ne j} \Theta(R_i + R_j - r_{ij})/N \simeq 0.526$ and $71.5\%$ of particles are not in contact in the region $0<x\leq 10$ and $|y| < R_{\rm tar}$, where $\Theta(x)$ and $N$ represent the Heaviside function and the number of particles in the region.

We evaluate physical quantities near the wall in two regions: $0<x\leq5d$ and $5d < x \leq 10d$, where we call (a) and (b) layers in the followings, respectively. We use $R_{\rm jet} /d = 15.0$ and $R_{\rm tar} /R_{\rm jet} = 2.2$ with the jet radius $R_{\rm jet}$.
We adopt the Cartesian coordinate, where $y=0$ is chosen to be the jet axis, 
and divide the calculation region into the $y$ direction $y = -5\Delta y, - 4 \Delta y, \cdots, 0, \cdots, 5\Delta y$, with $\Delta y \equiv R_{\rm tar} /5$.
Then we estimate physical quantities in the corresponding mesh region with $k\Delta y < y < (k+1)\Delta y$ ($k = -5, -4, \cdots,4$). Numerical data are averaged over ten initial configurations with the same $\tilde{\phi}_0$ and error bars in figures denote their variance.

We calculate the stress tensor as in Ref. \cite{stresstensor}. The microscopic definition of the stress tensor at {\bf r} is given by
\begin{equation}
\sigma_{\mu \nu}({\bf r}) = \frac{1}{A} \sum_{i} m u_{i \mu} u_{i \nu} + \frac{1}{A} \sum_{i<j} F_{\mu} ^{ij} r_{\nu} ^{ij},
\end{equation}
where $i$ and $j$ are indices of particles, $\mu, \nu = x,y$, the contact force between $i$ th and $j$ th particles $F^{ij} _{\mu}$ and $\sum_i$ denotes the summation over the particles denoted by $i$ located at ${\bf r}$. $A$ is the are of each mesh at ${\bf r}$ and $u_{i \mu} ({\bf r})= v_{\mu} ^{i} - {\bar v}_{\mu}({\bf r})$ with the mean velocity $ {\bar v}_{\mu}({\bf r})$ in the mesh at ${\bf r}$.

\section{Rheology of Granular Jets for the frictionless case}\label{rheol}

In this section, our numerical results of granular jet, for 2D frictionless cases are presented. The results for frictional grains will be reported in Sec. \ref{section_friction}. In Sec. 3.1, the existence of the dead zone and the profile of the area fraction are discussed. After showing profiles of the stress tensor in (a) or (b) layer in Sec. 3.2, we evaluate the equation of state and constitutive equation to compare our system with the critical behavior of the jamming in Secs. 3.3 and 3.4, respectively.

\begin{figure}[h]
\centering
\includegraphics[scale = 1.2]{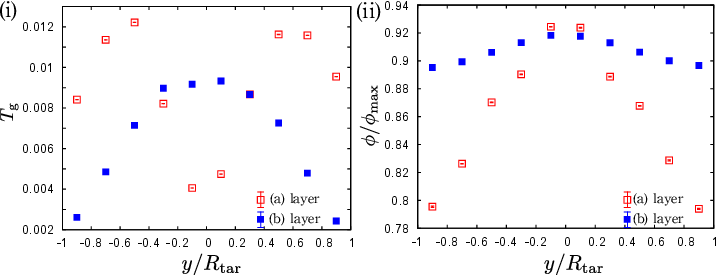}
\caption{The profile of $T_{\rm g}$ and $\phi / \phi_{\rm max}$ for the frictionless case with $\tilde{\phi}_0 = 0.90$ is shown in (i) and (ii), respectively.
There exists the dead zone in (a) layer, which is denoted by red empty squares. 
Blue filled squares denote $T_{\rm g}$ in (b) layer, where the dead zone does not exist. Grains are well packed in 2D: $0.79 < \phi/\phi_{\rm max} < 0.94$. Note that $\phi$ in (b) layer remains nearly constant compared with those in (a) layer, because grains in (b) layer are not compressed, while those in (a) layer are ejected after the compression.}
\label{dead}
\end{figure}

\subsection{Existence of the dead zone and the profile of the are fraction}
Chicago group suggested the existence of the dead zone near the target, where the motion of the grains is frozen, \cite{wendy, guttenberg}. 
Ellowitz et al. suggested that the dead zone exists in the sense that the velocity of grains are close to zero in Ref. \cite{wendy}. However, as is shown in our previous paper \cite{sano_hayakawa2}, although the velocity of grains at the center is small, the fluctuation of the velocity, i.e. the granular temperature $T_{\rm g}$, defined by $T_{\rm g} \equiv \sum_{i \in {\rm c}} m{\bf u}_i ^2 / DN_{\rm c} $ with the number of grains $N_{\rm c}$ in the mesh $c$ and the spatial dimensions $D$, is the largest at the center in 3D $(D = 3)$. 
 
On the other hand, we verify the existence of the actual frozen layer (a) i.e. $T_{\rm g} \simeq 0$. The fluctuation of the grain velocity in (a) layer is suppressed, while the motion is not frozen in (b) layer for 2D granular jets $(D = 2)$. 
The numerical data for $T_{\rm g}$ in 2D for the frictionless case are shown in Fig. \ref{dead} (i). $T_{\rm g}$ is the smallest at the center $y \simeq 0$ in (a) layer, which cannot be found in our previous 3D study (see Fig. 2 in Ref. \cite{sano_hayakawa2}), while $T_{\rm g}$ is the largest at $y \simeq 0$ in (b) layer. In very recent paper by Chicago group, it is suggested that the dead zone also exists in 3D experiment by introducing the effective temperature whose definition is not explicitly written \cite{wendy2}. \footnote{The differrence between our previous papers \cite{sano_hayakawa1, sano_hayakawa2} and their paper \cite{wendy2} might come from the difference of the jet size, in which they used $R_{\rm tar}/d \simeq 50.0$, but we used $R_{\rm tar}/d = 5.0$. }

The profile of the packing fractions divided by $\phi_{\rm max}$ with $\phi_{\rm max} \equiv \pi/(2\sqrt{3})\simeq 0.907$ in 2D are shown in Fig. \ref{dead} (ii) for the frictionless case. In 3D, the packing fraction divided by $\phi_{\rm max}^{3D} \simeq 0.740$ ranges within $0.30 < \phi/\phi_{\rm max}^{3D} < 0.75$. Compared with 3D, grains in 2D are well packed: $0.79 < \phi/\phi_{\rm max} < 0.94$. Note that $\phi$ in (b) layer is almost independent of the position, while $\phi$ in (a) layer strongly depends on the position.

\subsection{Profile of the stress tensor}
\begin{figure}[h]
\centering
\includegraphics[scale = 1.2]{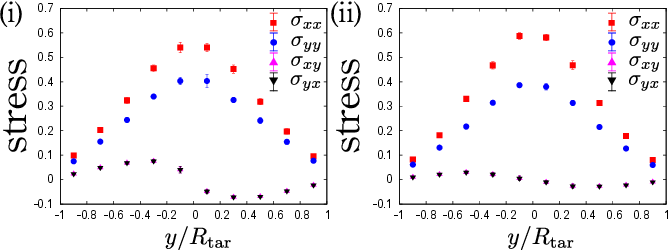}
\caption{ The profiles of the stress tensor in (a) and (b) layer for $\tilde{\phi}_0 = 0.90$ are shown in (i) and (ii), respectively. 
There exist large normal stress differences between $\sigma_{xx}$ and $\sigma_{yy}$, and the shear stress is much smaller than the normal stress in (b) layer, though it is not so small in (a) layer.}
\label{profile}
\end{figure}

The profiles of the stress tensor for (a) and (b) layers of frictionless grains are shown in Fig. \ref{profile} (i) and (ii), respectively. We stress that there exists a large normal stress difference between $\sigma_{xx}$ and $\sigma_{yy}$ in each layer as in 3D case \cite{sano_hayakawa1}.

Ellowitz {\itshape et al.} suggested that the profile of the velocity and the pressure for the granular jet are reproducible from the simulation of a perfect fluid \cite{wendy} but our result may not support their claim.
 Indeed, the shear stress looks small but finite. Moreover, the large normal stress difference exists in both layers, which does not exist in the perfect fluid. 
We should note that they have not discussed the stress tensor itself in details, though they reproduce some similar feature through their hard core simulation. In addition, Huang {\itshape et al.} indicated the relevant role of the contact stress in their DEM simulation, which may be an indirect objection to the perfect fluidity of the jet flow \cite{huang}.

\subsection{Equation of state}
Let us discuss the equation of state for the 2D granular jet impact. We estimate the strain rate $D_{xy} \equiv (\partial \bar{v}_y  / \partial x + \partial \bar{v}_x  / \partial y) /2$ as $\partial \bar{v}_y (\Delta x /2, y) / \partial x \simeq ( \bar{v}_y(3\Delta x /4, y) - \bar{v}_y(\Delta x /4, y)) / (\Delta x /2)$, $\partial \bar{v}_y (3\Delta x /2, y) / \partial x \simeq ( \bar{v}_y(7\Delta x /4, y) - \bar{v}_y(5\Delta x /4, y)) / (\Delta x /2)$ and  $\partial \bar{v}_x (x,y) / \partial y \simeq ( \bar{v}_x(x, y + \Delta y/2) - \bar{v}_x(x, y - \Delta y/2)) / \Delta y$. Since physical quantities are evaluated near the wall, the mesh $0 < x < \Delta x$ is divided into $0 < x \leq \Delta x /2$ and $\Delta x / 2 < x < \Delta x$, and $\Delta x \leq x < 2\Delta x$ is divided into $\Delta x \leq x < 3\Delta x /2$ and $3 \Delta x /2 \leq x < 2\Delta x$ to calculate $\partial \bar{v}_y (\Delta x/2, y) / \partial x$ and $\partial \bar{v}_y (3\Delta x /2, y) / \partial x$. $-R_{\rm tar}<y<R_{\rm tar}$ is divided into $-11 \Delta y /2 < y < -9 \Delta y/2, -9 \Delta y /2 < y < -7 \Delta y/2, \cdots 9 \Delta y /2 < y < 11 \Delta y/2$ to calculate $\partial \bar{v}_x (x, y) / \partial y$. 

\begin{figure}[h]
\centering
\includegraphics[scale = 1.2]{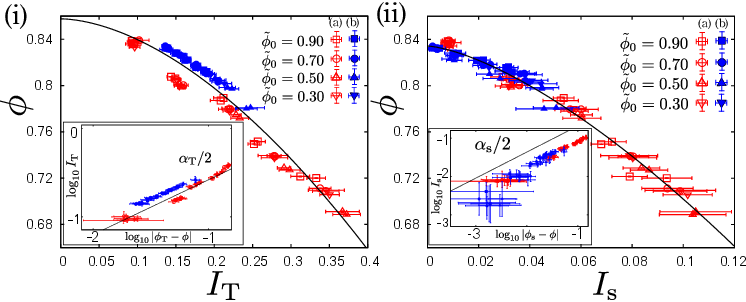}
\caption{Numerical data for the bi-disperse case of frictionless grains are plotted on $\phi$ vs $I_{\rm T}$ plane (i) and $\phi$ vs $I_{\rm s}$ plane (ii) of the main figure. Red and blue points denote data of (a) and (b) layer for several $\tilde{\phi}_0$, respectively. The corresponding solid lines in figures are fitting equations (\ref{fiteos1}) and (\ref{fiteos2}). The insets denote numerical data for (i) $\log_{10} I_{\rm T}\ {\rm vs}\ \log_{10} |\phi_{\rm T} - \phi|$ and (ii) $\log_{10} I_{\rm s}\ {\rm vs}\ \log_{10} |\phi_{\rm s} - \phi|$ to examine how good the fitting results are.}
\label{eos_fig}
\end{figure}

We follow the analysis in Ref. \cite{cruz}. Here, we introduce two dimensionless numbers consisting of pressure: $I_{\rm T} \equiv \sqrt{T_{\rm g} / Pd^2}$ and $I_{\rm s} \equiv D_{xy} \sqrt{m/P}$ with pressure $P \equiv (\sigma_{xx} + \sigma_{yy}) /2$. We plot numerical data on $\phi$ vs $I_{\rm T}$ plane and $\phi$ vs $I_{\rm s}$ plane, in Fig. \ref{eos_fig} (i) and (ii), respectively. Comparing $\phi$ in (a) with (b) layers against the identical $I_{\rm T}$, $\phi$ in (b) layer has a little larger value than $\phi$ in (a) layer at the same $I_{\rm T}$, while all $\phi$ against $I_{\rm s}$ are collapsed on a universal curve (Fig. \ref{eos_fig} (ii))

We can fit the data by the equations
\begin{eqnarray}
\phi &=& \phi_{\rm T} - a_{\rm T} I_{\rm T} ^{2 / \alpha_{\rm T}} \label{fiteos1}\\
\phi &=& \phi_{\rm s}- a_{\rm s} I_{\rm s} ^{2 / \alpha_{\rm s}}, \label{fiteos2}
\end{eqnarray}
with constants $\phi_{\rm T}, a_{\rm T}, \alpha_{\rm T}, \phi_{\rm s}, a_{\rm s}$ and $\alpha_{\rm s}$. Fitting parameters are determined simultaneously by using Levenberg-Marquardt algorithm \cite{fit}. The obtained equations of states are written as
\begin{eqnarray}
\frac{Pd^2}{T_{\rm g}} &=& \frac{a_{\rm T} ^{\alpha_{\rm T}}}{(\phi_{\rm T} - \phi) ^{\alpha_{\rm T}}}\label{eos1}\\ 
\frac{P}{mD_{xy} ^2} &=& \frac{a_{\rm s} ^{\alpha_{\rm s}}}{(\phi_{\rm s} - \phi) ^{\alpha_{\rm s}}}, \label{eos2}
\end{eqnarray}
where the comparison of Eqs. (\ref{fiteos1}) and (\ref{fiteos2}) with numerical data for the frictionless case are shown in the main figure of Fig. \ref{eos_fig} (i) and (ii), respectively. From Eqs. (\ref{fiteos1}) and (\ref{fiteos2}) which suggest the pressure diverging at $\phi_{\rm T}$ or $\phi_{\rm s}$ , the granular particles are well packed with the fraction sufficiently close to the jamming point. 

The obtained parameters from our simulation are $\phi_{\rm T} = 0.858 \pm 0.006, a_{\rm T} = 0.980 \pm 0.1, \alpha_{\rm T} = 1.15 \pm 0.1, \phi_{\rm s} = 0.834 \pm 0.001, a_{\rm s} = 3.94 \pm 0.5$ and $\alpha_{\rm s} = 1.36 \pm 0.05$, where the error originates from the fitting. We also plot $\log_{10} I_{\rm T}\ {\rm vs}\ \log_{10} |\phi_{\rm T} - \phi|$ and $\log_{10} I_{\rm s}\ {\rm vs}\ \log_{10} |\phi_{\rm s} - \phi|$ and the corresponding slope $\alpha_{\rm T}/2, \alpha_{\rm s}/2$ in the inset of Fig. \ref{eos_fig} (i) and (ii), respectively, to examine how good our fitting results are, by using obtained critical densities $\phi_{\rm T}$ and $\phi_{\rm s}$.
Note that the conventional jamming point $\phi_{\rm J} \simeq 0.8425$ at which the pressure diverges is located between $\phi_{\rm s}$ and $\phi_{\rm T}$ and close to $\phi_{\rm s}$ \cite{OH4}. 
The asymptotic divergences of pressure for the frictionless case are described as
\begin{equation}
\frac{Pd^2}{T_{\rm g}} \sim (\phi_{\rm T} - \phi) ^{- 1.15},\quad \frac{P}{mD_{xy} ^2} \sim (\phi_{\rm s} - \phi) ^{- 1.36}.
\end{equation}

In a conventional picture based on the extrapolation of the kinetic theory, the divergence of the pressure is expected to originate from the divergence of the radial distribution function i.e.
\begin{equation}
\frac{Pd^2}{T_{\rm g}} - 1 = \phi g(\phi) \propto \frac{\phi_c}{\phi_c - \phi} \label{rdf}
\end{equation}
as $\phi \to \phi_c$, with the radial distribution function $g(\phi)$
\begin{equation}
g(\phi)= \left\{
\begin{array}{ll}
\frac{1- 0.436\phi}{(1-\phi)^2} & \mbox{ ($0 < \phi < \phi_f $) } \\
 \frac{(1- 0.436\phi)(\phi_c - \phi_f)}{(1-\phi_f)^2(\phi_c - \phi)} & \mbox{ ($\phi_f < \phi < \phi_c $),}
\end{array}
\right. 
\end{equation}
the critical density $\phi_c = 0.82$ and the freezing density $\phi_f = 0.69$ \cite{torquato}. In our case, the data are not far from $Pd^2/T_{\rm g} \sim P/mD_{xy}^2 \sim (\phi_c - \phi)^{-1}$ expected from the conventional view based on the extrapolation of the kinetic theory, where $T_{\rm g} \sim md^2 D_{xy}^2$ is assumed. 

On the other hand, Hatano demonstrated an elegant scaling law in the vicinity of $\phi_{\rm J}$, where the corresponding exponents are estimated as $\alpha_{\rm s} = 2.8$ and $\alpha_{\rm T} = 1.7$ from his data of the jamming transition \cite{hatano_jam1}. Otsuki and Hayakawa showed the phenomenological explanation of the critical behavior near $\phi_{\rm J}$ and they predicted $\alpha_{\rm s} = 4.0$ and $\alpha_{\rm T} = 2.0$ \cite{OHL, OH3}. 
It should be stressed that the critical scaling of the jamming transition is analyzed in the $D_{xy} \to 0$ limit, and the critical exponents strongly depend on the choice of the jamming point.
Because the strain rate cannot be controlled in our setup, the jamming point is not clearly defined. Moreover, there are no data above the jamming transition in which the residual stress exists. Thus, our obtained exponents are smaller than those of the jamming transition for sheared granular particles.
We note that the data for $I_{\rm T}$ in (a) and (b) layers are separated, due to the difference of the profile of $T_{\rm g}$. 

\begin{figure}[h]
\centering
\includegraphics[scale = 0.8]{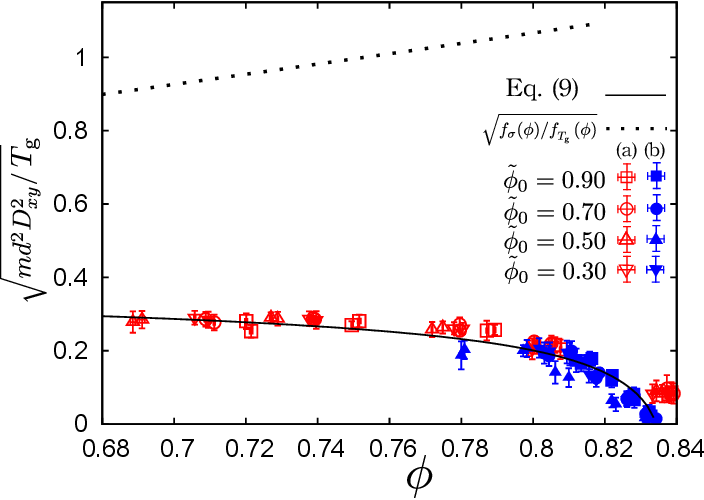}
\caption{The numerical data for $\sqrt{md^2 D_{xy} / T_{\rm g} }$ with Eq. (\ref{tg_dxy}) are compared for $\phi < \phi_{\rm s}$. The deviated data exist around $\phi \simeq 0.84$ in (a) layer, which may result from the existence of the source point at $y \simeq 0$. We also plot $\sqrt{f_{\sigma}(\phi)/f_{T_{\rm g}}(\phi)}$ as the dashed line for comparison.}
\label{eos_fig2}
\end{figure}

From Eqs. (\ref{eos1}) and (\ref{eos2}), $T_{\rm g}$ and $D_{xy}$ are expected to satisfy
\begin{equation}
\frac{md^2 D_{xy} ^2}{T_{\rm g}} = \frac{a_{\rm T} ^{\alpha_{\rm T}} (\phi_{\rm s} - \phi)^{\alpha_{\rm s}} }{a_{\rm s} ^{\alpha_{\rm s}} (\phi_{\rm T} - \phi)^{\alpha_{\rm T}}}\label{tg_dxy}.
\end{equation}
The validity of Eq. (\ref{tg_dxy}) is verified in Fig. \ref{eos_fig2}, which can be independent check of the scaling laws (\ref{eos1}) and (\ref{eos2}). From Fig. \ref{eos_fig2}, Eq. (\ref{tg_dxy}) well reproduces the data for $\phi < \phi_{\rm s}$. Numerical data for $\phi \simeq 0.84$ around the center, deviates from Eq. (\ref{tg_dxy}), which may result from the existence of the source point. The velocity field at the center is singular, compared with other regions. 
Actually, the similar deviation of the numerical data at the center from the theory can be found in our previous 3D study, in terms of the pressure and the shear viscosity \cite{sano_hayakawa1}. 

The relation $T_{\rm g} \propto md^2D_{xy}^2$, which is equivalent to the Bagnold's scaling, is known to be derived from the energy balance equation for dense granular flow in the case that the heat flux can be negligible \cite{incl1}. The dotted line in Fig. \ref{eos_fig2} represents the curve which can be derived from Ref. \cite{jenkins_richman} by taking the frictionless limit, which is written as $md^2 D_{xy} / T_{\rm g} = f_{\sigma}(\phi)/f_{T_{\rm g}}(\phi)$ with
\begin{eqnarray}
f_{\sigma}(\phi) &\equiv& f_{\sigma} ^{(0)}(\phi) + \frac{2(1+e)}{\pi ^{3/2}}g(\phi),\\
f_{T_{\rm g}}(\phi) &\equiv& \frac{16}{\pi ^{3/2}} (1-e^2)\phi^2 g(\phi),\\
f_{\sigma} ^{(0)}(\phi) &\equiv& \frac{2}{\sqrt{\pi}(7-3e) g(\phi)} \left\{ 1 + \frac{1+e}{2}\phi g(\phi)\right\} \left\{1 + \frac{(3e-1)(1+e)}{4}\phi g(\phi) \right\}.
\end{eqnarray}
Because there exists the unique critical density $\phi_c$ for $f_{\sigma}(\phi)$ and $f_{T_{\rm g}}(\phi)$, the conventional curve does not exhibit the critical behavior:
\begin{equation}
\frac{f_{\sigma}(\phi)}{f_{T_{\rm g}}(\phi)} \to \frac{64(7-3e)(1-e)\phi_c ^2}{{\phi_c ^2 \pi}(3e-1)(1+e) + 8(7-3e)}
\end{equation}
as $\phi \to \phi_c$, while $md^2D_{xy}/T_{\rm g} \to 0$ as $\phi \to \phi_{\rm s}$ in our setup, due to the two critical densities $\phi_{\rm s} < \phi_{\rm T}$. It should be noted that the functional form of $f_{\sigma}, f_{\sigma} ^{(0)}$ and $f_{T_{\rm g}}$ vary, depending on the level of approximation.

Although there exist the inhomogeneity of $T_{\rm g}$, as in the dead zone near the target. This is because we can generalize the discussion of Bagnold's scaling, at least, if the inhomogeneity is small (see Appendix B). Indeed, our case satisfies the condition that the gradient of  the thermal velocity $\sqrt{2T_{\rm g}/m}$ is much smaller than that of the velocity field. Because the momentum transfer plays major roles in the energy balance equation, where the only relevant time scale would be the shear rate. The detail analysis for the inhomogeneity of $T_{\rm g}$ is shown in Appendix B.

\subsection{Constitutive equation}
\subsubsection{Effective friction constant}
\begin{figure}[h]
\centering
\includegraphics[scale = 0.65]{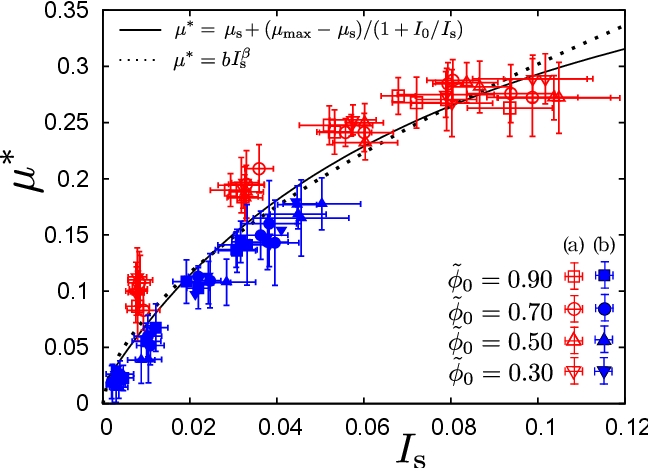}
\caption{Numerical data for the bi-dispersed frictionless grains are plotted on $\mu^*$ vs $I_{\rm s}$ plane. Red and blue points denote data for (a) and (b) layer for several $\tilde{\phi}_0$, respectively. All points are fitted into the phenomenological equation in Eqs. (\ref{conv}) and (\ref{pl}), where we cannot judge which equation is better from the data.}
\label{friction_bi}
\end{figure}

Let us discuss $I_{\rm s}$ dependence of effective friction constant $\mu^* \equiv -\sigma_{xy} /P$ to obtain the constitutive equation. 
Numerical data for the frictionless case are shown in Fig. \ref{friction_bi}. 
The behavior of $\mu^*$ is conventionally described as 
\begin{equation}
\mu^* (I_{\rm s}) = \mu_{\rm s} + \frac{\mu_{\rm max} - \mu_{\rm s}}{1 + I_0 / I_{\rm s}}, \label{conv}
\end{equation}
where $\mu^*$ starts from a static value of $\mu_{\rm s}$ at zero shear rate and converges to a limiting value of $\mu_{\rm max}$ at high $I_{\rm s}$. We obtain $\mu_{\rm s} = 0.0153 \pm 0.009, \mu_{\rm max} = 0.521 \pm 0.06$ and $I_0 = 0.0820 \pm 0.02$ for the frictionless case
by fitting. Thus, $\mu^* (I_{\rm s})$ can be fitted by the conventional relation (\ref{conv}), which is denoted by a solid line in Fig. \ref{friction_bi}.
It should be stressed that $\mu_{\rm s}$ is close to zero. 

Some researchers proposed a different constitutive equation called the power-law friction
\begin{equation}
\mu^* = \mu_{\rm s} + b I_{\rm s} ^{\beta}, \label{pl}
\end{equation}
which also well reproduces numerical data \cite{GDR, cruz, hatano,poliq,hatano_jps, ohern}, where $\beta$ ranges from $0.28$ to $1.0$, depending on the dimension, microscopic parameters and the friction constant of grains. Numerical data can be fitted by Eq. (\ref{pl}) within error bars, where we obtain $b = 1.18 \pm 0.1$ and $\beta = 0.592 \pm 0.03$, assuming $\mu_{\rm s} = 0$ for the frictionless case. The fitting result of Eq. (\ref{pl}) is denoted by a dotted line in Fig. \ref{friction_bi}. As can be seen in Fig. \ref{friction_bi}, there is no significant difference between Eqs. (\ref{conv}) and (\ref{pl}). We, of course, cannot discuss the superiority of one of frictional laws from our simulation. We should stress that the fitting for both Eqs. (\ref{conv}) and (\ref{pl}) leads to almost zero $\mu_{\rm s}$. This implies that the residual stress is negligible in the granular fluid after the jet impact.

\subsubsection{The asymptotic divergence of the shear stress}
\begin{figure}
\centering
\includegraphics[scale = 1.4]{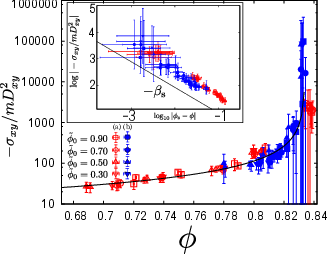}
\caption{Equation (\ref{shear_st}), which denotes the divergence of the shear stress, is independently checked for the bi-disperse frictionless case. Red and blue points denote data for (a) and (b) layer for several $\tilde{\phi}_0$, respectively. Equation (\ref{shear_st}) well reproduces numerical results. Because $\sigma_{xy}$ itself in (b) layer are small, the error bars in (b) layer are larger than those in (a) layer. In the inset, we plot $\log_{10} |-\sigma_{xy}/mD_{xy}^2|$ vs $\log_{10} |\phi_{\rm s} - \phi|$, to examine how good Eq. (\ref{shear_st}) is.}
\label{shear_div}
\end{figure}

Let us discuss the asymptotic divergence of the shear stress:
\begin{equation}
-\frac{\sigma_{xy}}{mD_{xy} ^2} \sim (\phi_{\rm s} - \phi)^{-\beta_{\rm s}}
\end{equation}
with an exponent $\beta_{\rm s}$.
By using the divergence of the pressure (\ref{eos2}) and the power-law friction $\mu^* \propto I_{\rm s} ^{\beta}$, we obtain the constitutive equation for $\sigma_{xy}$
\begin{equation}
-\frac{\sigma_{xy}}{mD_{xy} ^2} = \frac{b a_{\rm s} ^{(1 - \beta /2)\alpha_{\rm s} }}{(\phi_{\rm s} - \phi)^{(1 - \beta /2)\alpha_{\rm s}}}, \label{shear_st}
\end{equation}
which is checked independently against the numerical data (Fig. \ref{shear_div}). 
The exponent is estimated to be $\beta_{\rm s} = (1 - \beta /2)\alpha_{\rm s} \simeq  0.96$. We also plot $\log_{10}  |\phi_{\rm s} - \phi| \ {\rm vs}\ \log_{10} |-\sigma_{xy}/mD_{xy}^2|$ with obtained $\phi_{\rm s}$ and the slope $-\beta_{\rm s}$ in the inset of Fig. \ref{shear_div}. The numerical data for large $\phi$ are deviated from the theoretical curve, due to the small shear stress and shear rate at $y \simeq 0$.
Because we use the power-law friction $\beta > 0$, the divergence of shear stress may be slightly weaker than that of $P$. 
Here, Bagnold's scaling is still satisfied even in the vicinity of the ``jammed" density, which is in contrast to the  actual jamming transition \cite{liu_nagel, OH1, Teitel, OHL, hatano_jam,OH2, OH3, garcia, losert,OH4,Teitel0,tighe,nord,hatano_jam1}.

Let us compare our observed critical behavior of the shear stress with the case of the jamming transitions, in details as well as that of the extrapolation of the kinetic theory. When we adopt the extrapolation of the kinetic theory, we have $-\sigma_{xy}/mD_{xy} ^2 = \eta /mD_{xy} \sim \eta^* \sim \phi^2 g(\phi) \sim \phi_c ^2/(\phi_c - \phi)$ as $\phi \to \phi_c$, i.e. $\beta_{\rm s} = 1.0$, where $T_{\rm g} \sim md^2 D_{xy}^2$ is used and dimensionless shear viscosity $\eta^* \equiv \eta / \eta_0$ is introduced with $\eta_0 \equiv \sqrt{m T_{\rm g} /4\pi d^2}$ and shear viscosity $\eta \equiv - \sigma_{xy} / D_{xy}$. 
The extrapolation from the kinetic regime by Garcia-Rojo {\itshape et al.} predicts that $\sigma_{xy}$ diverges at density different from $Pd^2/T_{\rm g}$ and $\beta_{\rm s} = 1.0$ \cite{garcia}.
Therefore, our analysis based on the power-law friction Eq. (\ref{pl}) predicts the results similar to Garcia-Rojo {\itshape et al} \cite{garcia}: $\beta_{\rm s} = 1.0$ and $\phi_{\rm s} < \phi_{\rm T}$.
On the other hand, the exponents for the divergence of $-\sigma_{xy}/mD_{xy}^2$ at the jamming transition for sheared granular materials are estimated to be $\beta_{\rm s} \simeq 2.6$ from data in Ref. \cite{hatano_jam1} and $\beta_{\rm s} = 4.0$ in Ref. \cite{OHL}. Thus, our corresponding exponent $\beta_{\rm s} = 0.96$ is much smaller than those of the jamming transition for sheared granular systems and rather close to the result of the kinetic theory. Otsuki {\itshape et al}. \cite{OHL} studied the difference of soft core jamming and the asymptotic divergence of hard core systems. Then, they confermed the exponent $\beta_{\rm s}$ can only deviate from 1.0 in very narrow critical region, in which the soft core effect becomes relevant. Although our system has high density, the number of particles in contacts is still not large. Therefore, we may regard the granular fluid after the impact as a hard core fluid.

\section{Effect of the friction constant}\label{section_friction}
\begin{figure}[h]
\centering
\includegraphics[scale = 2.5]{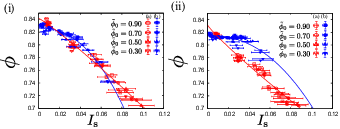}
\caption{Fitting results of Eq. (\ref{fiteos2}) for $\mu_{\rm p} = 0.2$ (i) and $\mu_{\rm p} = 1.0$ (ii). As $\mu_{\rm p}$ becomes larger, data for (a) and (b) layer deviates from each other. }
\label{frictional_bagnold}
\end{figure}

\begin{figure}[h]
\centering
\includegraphics[scale = 1.2]{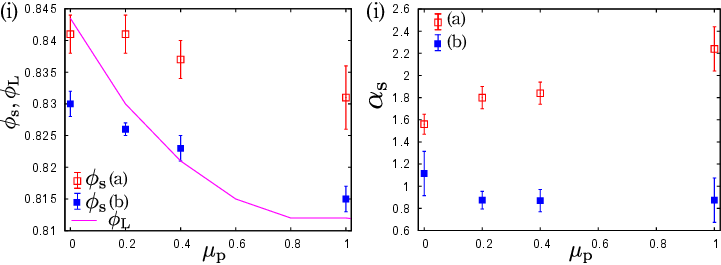}
\caption{The $\mu_{\rm p}$ dependence of the critical density $\phi_{\rm s}$ and the exponent $\alpha_{\rm s}$ in (i) and (ii), respectively. $\phi_{\rm s}$ in both (a) and (b) layer slightly decrease as $\mu_{\rm p}$ increases. The purple solid line denotes the corresponding critical density of jamming for frictional granular particles $\phi_{\rm L}$ \cite{OH1}. The exponents in (a) layer increases as $\mu_{\rm p}$ increases, while they decrease in (b) layer.}
\label{friction_density}
\end{figure}

We examine how the fluid state depends on the friction constant from the simulation for $\mu_{\rm p} = 0.2, 0.4$ and $1.0$. 
It is noteworthy that the separation between (a) and (b) layer exists for larger $\mu_{\rm p}$, even on $\phi$ vs $I_{\rm s}$ plane. The results for $\mu_{\rm p} = 0.2$ and $\mu_{\rm p} = 1.0$ are shown in Fig. \ref{frictional_bagnold} (i) and (ii), respectively. Because there are two branches on $\phi$ vs $I_{\rm s}$ plane, we adopt Eq. (\ref{fiteos2}) to fit the data in (a) or (b) layer, separately. Figure \ref{friction_density} denotes the critical densities and the exponents for each $\mu_{\rm p}$, where $\phi_{\rm s}$ in both (a) and (b) layer slightly decrease as $\mu_{\rm p}$ increases, and $\alpha_{\rm s}$ in (a) layer increases as $\mu_{\rm p}$ increases, while it decreases in (b) layer. We note that the decrease of our critical densities $\phi_{\rm s}$ both in (a) and (b) layer are gentler than $\phi_{\rm L}$ of the jamming transition for sheared granular systems \cite{OH1}.


\begin{figure}[h]
\centering
\includegraphics[scale = 1.8]{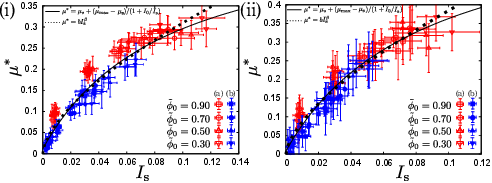}
\caption{Numerical data for $\mu_{\rm p} = 0.2$ (i) and $\mu_{\rm p} = 1.0$ (ii) can be fitted into Eqs. (\ref{conv}) and (\ref{pl}) within error bars, where we cannot judge which equations are better. The friction law is little affected by the friction of grains. It should be noted that $\mu^*$ monotonically increases from near zero, as the increment of $I_{\rm s}$, even for large $\mu_{\rm p}$.}
\label{frictional_friclaw}
\end{figure}

In contrast, the friction law is little affected by the friction of grains. The results for the friction law are shown in Fig. \ref{frictional_friclaw} (i) for $\mu_{\rm p} = 0.2$ and (ii) for $\mu_{\rm p} = 1.0$, where the numerical data can be fitted by both Eqs. (\ref{conv}) and (\ref{pl}). We stress that $\mu^*$ monotonically increases from near zero, as $I_{\rm s}$ increases, even for large $\mu_{\rm p}$. 

\section{Result for the mono-disperse case}\label{discussion}

\begin{figure}[h]
\centering
\includegraphics[scale = 0.4]{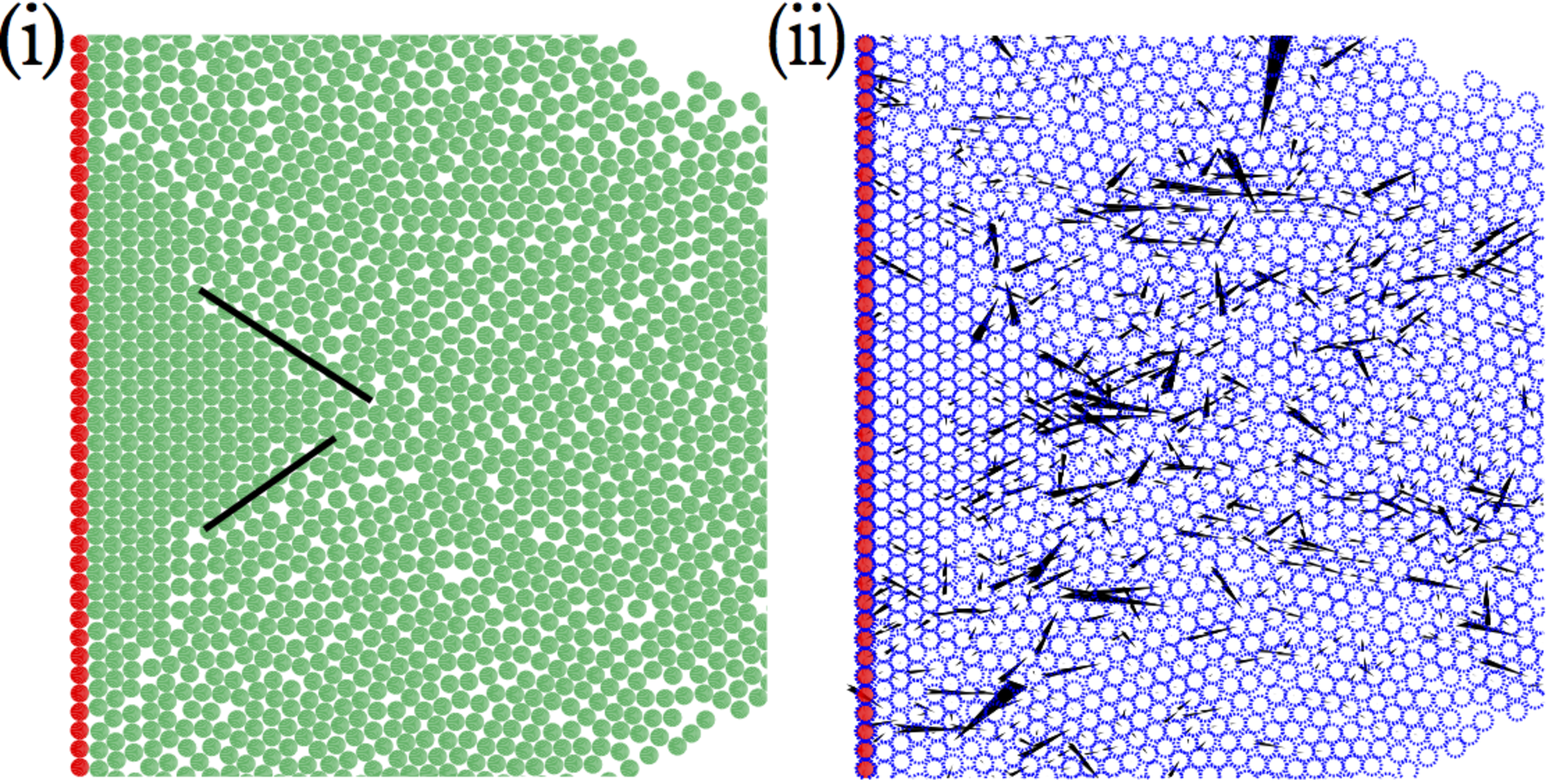}
\caption{Typical snapshot of the simulation for the frictionless and mono-disperse grains case with $\tilde{\phi}_0 = 0.90$ near the target. Green particles denote mobile grains in the granular jet and red particles are wall-particles (i). The black solid lines are drawn by hand to clarify the grain boundary between the crystallized region and the disordered region, where the boundary becomes a slip line. Crystallization into a triangular lattice can be seen near the region enclosed by the black lines. All of the corresponding contact forces between grains are visualized as black colored arrows in (ii).}
\label{snap_mono}
\end{figure}

Here, we discuss the impact of  granular jets in 2D for the mono-disperse case. A typical snapshot zoomed near the target is shown in Fig. \ref{snap_mono}, where grains are crystalized near the wall. The black solid lines in Fig. \ref{snap_mono} (i) are drawn by hand to clarify the grain boundary between the crystallized region and the disordered region, where the boundary becomes a slip line. We also visualize all of the corresponding contact force network in Fig. \ref{snap_mono} (ii).

\begin{figure}[h]
\centering
\includegraphics[scale = 1.23]{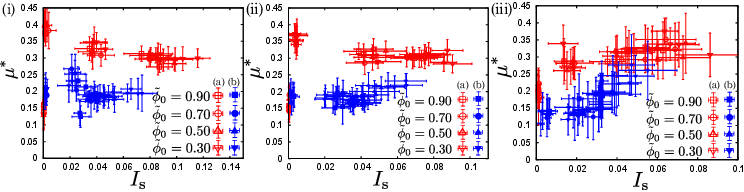}
\caption{Numerical data for the frictionless case (i), $\mu_{\rm p} = 0.2$ (ii) and $\mu_{\rm p} = 1.0$ (iii) are plotted on $\mu^*$ vs $I_{\rm s}$ plane. 
Red and blue points denote data for (a) and (b) layer for several $\tilde{\phi}_0$, respectively. Unlike bi-disperse cases, $\mu^* (I_{\rm s})$ for (a) and (b) cannot be fitted into a single curve. $\mu^* (I_{\rm s})$ in (i)-(iii) show similar behavior in (a) layer. However, in (b) layer, because frictionless grains cannot roll over the crystallized region, $\mu^* (I_{\rm s})$ for the frictionless case shows similar dependence on $I_{\rm s}$ to that in (a) layer, while the corresponding $\mu^* (I_{\rm s})$ for frictional cases do not.}
\label{friction_mono}
\end{figure}

The notable difference of the mono-disperse cases from the bi-disperse cases appears in the friction law. We plot $\mu^* (I_{\rm s})$ for the mono-disperse case of frictionless grains, $\mu_{\rm p} = 0.2$ and $\mu_{\rm p} = 1.0$ in Fig. \ref{friction_mono} (i), (ii) and (iii), respectively. First of all, $\mu ^* (I_{\rm s})$ for (a) layer and (b) layer cannot be fitted by a single curve, unlike the bi-disperse case. Judging from the snapshot (Fig. \ref{snap_mono}), grains, at least, in (a) layer are partially crystallized. Therefore, it is reasonable that the response of the crystallized region is different from that in disordered regions in (b) layer. 

The behavior of $\mu^* (I_{\rm s})$ in (a) layer, which are observed in both frictional and frictionless cases, can be understood as follows. 
Because of the crystallization, a grain is trapped in a crystallized region. However, as $I_{\rm s}$ increases, the grain can escape from the crystallized region. Thus, $\mu^*$ decreases as $I_{\rm s}$ increases. 

The macroscopic friction $\mu^*(I_{\rm s})$ for frictionless grains are different from that for frictional grains in (b) layer. 
The most remarkable difference between the frictionless and the frictional cases is the existence of peak of $\mu^*$ at a small $I_{\rm s}$ for the frictionless case, while there is no such a peak for frictional cases. 
Because a frictional grain can roll over grains, grains easily form a cluster. 
Therefore, the boundary between such clusters becomes a slip line. 
Thus, $\mu^* (I_{\rm s})$ would be constant as $I_{\rm s}$ becomes smaller. On the other hand, because a frictionless grain can neither roll over them nor slip,  it is trapped in the crystallized region even for the large $I_{\rm s}$. Thus, $\mu^*(I_{\rm s})$ for frictional and frictionless cases exhibit different behaviors in (b) layer. However, we should stress that there exist two metastable branches for both frictionless and frictional cases.


\section{Discussion and Conclusion}\label{concl}

We have performed two-dimensional simulations for the impact of a granular jet and discussed its rheology. 
We confirmed the existence of the dead zone, as is reported  in Ref. \cite{wendy} at least in (a) layer, unlike our previous three-dimensional cases \cite{sano_hayakawa1, sano_hayakawa2}. 
There exists large normal stress difference, which has not been reported previously. 
The shear stress is much smaller than the normal stress, at least in (b) layer. We need to solve the inconsistency in (a) layer with Ref. \cite{wendy}.

We have analyzed the rheology of frictionless grains after the jet impact.
We found that the pressure and the shear stress diverge with exponents similar to the extrapolations from the kinetic regime, and their exponents are smaller than those of the jamming transition for sheared granular systems. We adopted the power-law friction for $\mu^*$ Eq. (\ref{pl}) to obtain the critical exponent for $\sigma_{xy}/mD_{xy}^2$. The discrepancy between our case and the jamming transition for sheared granular systems would originates from (i) our system cannot reach the true jamming transition and (ii) the uncontrollability of $D_{xy}$ in our setup. The jamming point for a sheared system $\phi_{\rm J}$ is located between $\phi_{\rm s} < \phi_{\rm J} < \phi_{\rm T}$ and is close to $\phi_{\rm s}$
Our analysis based on the power-law friction is consistent with that by Garcia-Rojo {\itshape et al} \cite{garcia}, where $\sigma_{xy}$ diverges at the density different from $Pd^2/T_{\rm g}$ \cite{garcia}. 

The effects of the friction of grains $\mu_{\rm p}$ have been discussed. Although Guttenberg \cite{guttenberg} suggested that $\mu_{\rm p}$ does not play a significant role, at least, in the scattering angle via the approximate hard-sphere method \cite{hybrid}, we found that the existence of the friction affects rheology of granular fluids after the impact. The separation between (a) and (b) layer appears for larger $\mu_{\rm p}$, even on $\phi$ vs $I_{\rm s}$ plane. The critical fraction $\phi_{\rm s}$ decreases as $\mu_{\rm p}$ increases, which is similar to the behavior of critical fraction of jammed frictional grains $\phi_{\rm L}$. The corresponding exponent $\alpha_{\rm s}$ increases (decreases) as the increment of $\mu_{\rm p}$ in (a) layer ((b) layer). 

The effective friction constant $\mu^* (I_{\rm s})$ for the mono-disperse case has two branches because of the coexistence  of the crystallized state and a liquid state. On the other hand, $\mu^* (I_{\rm s})$ for the bi-disperse case can be described by known constitutive equations for dense granular flow\cite{jop,poliq,GDR,cruz,hatano,hatano_jps,ohern,hill}. 

Finally, let us comment on the rheological model proposed in a recent paper of Chicago group \cite{wendy2}. It is suggested that the granular fluid after the impact may be described by the plastic flow without the viscous stress and with the isotropic pressure. This suggestion is interesting, but our data may not support their suggestion. In fact, our data suggest the existence of viscous term (the shear stress depends on the location), the pressure is anisotropic, and no evidence of the existence of the residual stress as is shown in Fig. \ref{friction_bi}. 

\section*{Acknowledgment}
We thank M. Otsuki for valuable discussions. A part of numerical computation in this work was carried out at the Yukawa Institute Computer Facility. This work is partially supported by the Grant-in-Aid for the Global COE program gThe Next Generation of Physics, Spun from Universality and Emergence hfrom MEXT, Japan and Grant-in-Aid for Scientific Research from MEXT (No. 25287098).

\appendix
\def\thesection{Appendix \Alph{section}}
\section{On artificial burst-like flows for the large $\mu_{\rm p}$ case}
\begin{figure}[h]
\centering
\includegraphics[scale = 1.0]{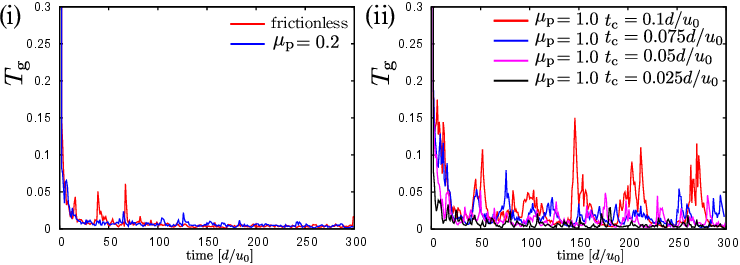}
\caption{The time evolution of $T_{\rm g}$ at $-\Delta y<y<0$ in (a) layer for the frictionless and $\mu_{\rm p} = 0.2$ (i) and $\mu_{\rm p} = 1.0$ for the several stiffness of grains (ii).  As $t_{\rm c}$ becomes smaller, i.e. as grains become stiffer, the burst-like flows are suppressed.}
\label{t_evol}
\end{figure}

In this appendix, we comment on the artificial burst-like flow in 2D, which appears in the case of large $\mu_{\rm p}$ with softer grains than those in the text. After the impact of a jet composed of softer grains with large $\mu_{\rm p}$, the burst occurs when a grain slips, because large tangential force can be accumulated before the slip of a grain. In Fig. \ref{t_evol}, we show the time evolutions of $T_{\rm g}$ at $-\Delta y<y<0$ in (a) layer for (i) the frictionless case and $\mu_{\rm p} = 0.2$, and (ii) $\mu_{\rm p} = 1.0$ with several stiffness, where $T_{\rm g}$ for the frictionless case and $\mu_{\rm p} = 0.2$ reaches the small steady values, while $T_{\rm g}$ raise many times after the impact for $\mu_{\rm p} = 1.0$ with large $t_{\rm c}$, due to the slip events. As $t_{\rm c}$ becomes smaller, the burst-like flows are suppressed. Thus, we use harder grains for large $\mu_{\rm p}$. Though there are a few small raises of $T_{\rm g}$ for the frictionless case, they are out of our averaging time. 

\section{Inhomogeneity of $T_{\rm g}$}
Here, let us discuss the effect of the inhomogeneity of $T_{\rm g}$ to Bagnold's scaling. We demonstrate that $T_{\rm g} \sim md^2D_{xy}^2$ may be valid because the gradient of $\sqrt{2T_{\rm g}/m}$ is much smaller than that of the velocity field in our setup.

\begin{figure}[h]
\centering
\includegraphics[scale = 1.0]{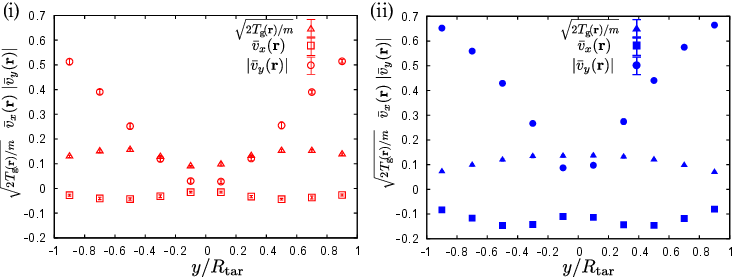}
\caption{Red empty (i) and blue filled (ii) points denote the data in (a) and (b) layer, respectively. The corresponding triangle, square and circle points are the data for $\sqrt{2T_{\rm g}/m}$, $\bar{v}_x$ and $|\bar{v}_y|$, respectively.  Although $\sqrt{2T_{\rm g}/m}$ is the smallest at the center $y \simeq 0$ in (a) layer, the inhomogeneity of $T_{\rm g}$ is smaller than that of $\bar{v}_x$ and $|\bar{v}_y|$. $|v_y|$ linearly increases as $y \to R_{\rm tar}$ from zero.}
\label{fig_appb}
\end{figure}

The numerical data for the profile of $\sqrt{2T_{\rm g}/m}$, $\bar{v}_{x}$ and $|\bar{v}_{y}|$ are shown in Fig. \ref{fig_appb}(i)(ii). Red empty (i) and blue filled (ii) points denote the data in (a) and (b) layer, respectively. The corresponding triangle, square and circle points are the data for $\sqrt{2T_{\rm g}/m}$, $\bar{v}_x$ and $|\bar{v}_y|$, respectively in Fig. \ref{fig_appb}. Although $\sqrt{2T_{\rm g}/m}$ is the smallest at the center $y \simeq 0$ in (a) layer, the inhomogeneity of $T_{\rm g}$ is much smaller than that of $\bar{v}_x$ and $|\bar{v}_y|$. In particular, it is notable that $|v_y|$ linearly increases as $|y| \to R_{\rm tar}$ from zero.

The relation $T_{\rm g} \sim md^2D_{xy}^2$ can be derived from the energy balance equation in the case that the heat flux can be negligible:
\begin{equation}
\sigma_{xy}D_{xy} = - \Gamma,
\end{equation}
where $\Gamma(\phi, T_{\rm g}) = (1-e^2)\tilde{\Gamma}(\phi) T_{\rm g} ^{3/2}/(m^{1/2}d^4)$ denotes the energy dissipation rate with dimensionless one $\tilde{\Gamma}(\phi)$. From $\sigma_{xy} = - \eta D_{xy}  = -\tilde{\eta}(\phi)m^{1/2}T_{\rm g} ^{1/2} D_{xy}/d^2$, we obtain $T_{\rm g} = \tilde{\eta}md^2D_{xy} ^2 / \{(1-e)^2 \tilde{\Gamma} \} \propto md^2 D_{xy}^2$. 

This relation, which is the basis of Bagnold's scaling, is unchanged even if we introduce small inhomogeneity for the density and the temperature. Indeed, we expand $\phi$ and $T_{\rm g}$ around homogeneous value:
\begin{eqnarray}
\phi &\simeq& \phi_0 + \delta \phi(x,y)\\
T_{\rm g} &\simeq& T_0 + \delta T(x,y),
\end{eqnarray}
where
\begin{equation}
T_0 \equiv \frac{\tilde{\eta}(\phi_0)md^2 D_{xy} ^2}{(1-e^2)\tilde{\Gamma}(\phi_0)}
\end{equation}
holds. Energy balance equation for the inhomogeneous case would be expressed as
\begin{equation}
-\eta D_{xy} ^2 (x,y) - \nabla \cdot \kappa(\phi, T_{\rm g}) \nabla T_{\rm g} = - \Gamma \label{inh_en},
\end{equation}
with the thermal conductivity $\kappa = \tilde{\kappa}(\phi) (T_{\rm g} /m)^{1/2}/d^2$ and dimensionless one $\tilde{\kappa}$, where we ignore the small density diffusive term which becomes zero in the elastic collisions.
By linearizing Eq. (\ref{inh_en}), $\delta T$ satisfies 
\begin{eqnarray}
\left(\nabla^2 - k^2 \right) \delta T&=&  -\frac{1}{d^2}\tilde{T}(x,y), \label{linearized_eq}\\
\tilde{T}(x,y) &\equiv& -\frac{\tilde{\eta}(\phi_0)}{\tilde{\kappa}(\phi_0)}\left\{ \frac{ \tilde{\Gamma}'(\phi_0)}{{\tilde{\Gamma}(\phi_0)}} - \frac{\tilde{\eta}'(\phi_0)}{\tilde{\eta}(\phi_0)} \right\}\delta \phi(x,y) md^2D_{xy}^2(x,y)
\end{eqnarray}
with $k \equiv \sqrt{(1-e^2) \tilde{\Gamma}(\phi_0)/d^2}$, $\tilde{\eta}(\phi) \equiv f_{\sigma}(\phi), \tilde{\Gamma}(\phi) = f_{T_{\rm g}}(\phi)$ and
\begin{equation}
\tilde{\kappa}(\phi) \equiv \frac{16}{(1+e)(19-15e)\sqrt{\pi}}\left(1+\frac{3}{8}\phi g(\phi)(1+e)^2 (2e-1) \right),
\end{equation}
where we use the frictionless limit of the results in Ref. \cite{jenkins_richman}.
We note that $\tilde{T}(x,y) \propto md^2 D_{xy} ^2$ holds. By introducing Green's function $G(x,y| x',y')$ we solve the inhomogeneous modified Helmholtz eq. (\ref{linearized_eq}) under Dirichlet condition $\delta T_{\rm C}(y) \equiv \delta T(x = 0,y)$ in the half space $D = \{(x,y)| x > 0\}$ with the boundary $C = \{(x,y)| x = 0 \}$. Green's function for modified Helmhotlz eq. in 2D for infinite space is given by
\begin{eqnarray}
{\mathcal L} G_{\rm free} (x,y| x',y')&=& - \delta(x - x')\delta(y-y')\\
G_{\rm free}(x,y| x',y')&=& \frac{1}{2\pi}K_0 (k \sqrt{(x - x')^2 + (y - y')^2}),
\end{eqnarray}
with the modified Helmhotlz operator ${\mathcal L} \equiv \nabla^2 - k^2$ and modified Bessel function for the second kind $K_0 (z) = \int_0 ^{\infty} dt e^{-z \cosh t} $. By Green's theorem,
\begin{eqnarray}
\int _D dxdy \left\{ G(x,y| x',y') {\mathcal L} u - u {\mathcal L}G(x,y| x',y') \right\} = \int_{-\infty} ^{\infty} dy'' \left(G\frac{\partial u}{\partial x''} - u \frac{\partial G}{\partial x''}\right) 
\end{eqnarray}
holds for an arbitrary function $u(x,y)$. Replacing $u = \delta T(x,y)$ and adopting $G(x = 0, y| x', y') = 0$, we obtain
\begin{equation}
\delta T(x',y') = \int_D dxdy G(x,y|x',y') \frac{1}{d^2}\tilde{T}(x,y) - \int_{-\infty} ^{\infty} dy'' \delta T(x'' = 0, y'') \frac{\partial }{\partial x''} G(x'' = 0,y''| x',y'),
\end{equation}
where we used $G(x'' = 0,y''| x',y') = 0$. Thus, the solution for Eq. (\ref{inh_en}) under the Dirichlet condition is represented as
\begin{equation}
\delta T(x,y) = \int_D dx'dy' G(x',y'|x,y) \frac{1}{d^2}\tilde{T}(x',y') - \int_{-\infty} ^{\infty} dy'' \delta T_{\rm C}(y'') \frac{\partial }{\partial x''} G(x'' = 0,y''| x,y) \label{delta_t_cal}.
\end{equation}
\begin{figure}[h]
\centering
\includegraphics[scale = 1.2]{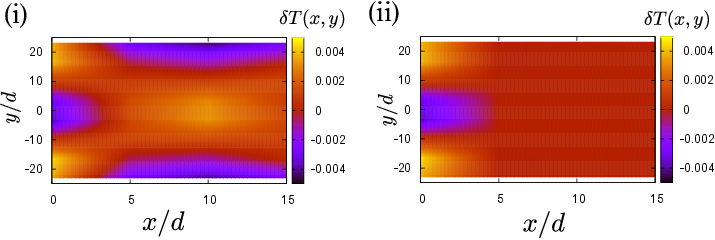}
\caption{Comparison between simulation data (i) and Eq. (\ref{delta_t_cal}) (ii). The inhomogeneity of $\delta T$ appeared in the dead zone rapidly decreases as $x \to \infty$. The characteristic length for the inhomogeneity is calculated to be $1/kd \simeq 0.696 < \Delta x$. }
\label{green_func_comp}
\end{figure}
Here, Green's function, which satisfies the condition $G(x = 0, y| x', y') = 0$ can be constructed as
\begin{eqnarray}
G(x,y|x',y') &=& G_{\rm free}(x,y| x',y') - G_{\rm free}(x,y| -x',y')\\
&=& \frac{1}{2\pi} \left\{K_0 (k \sqrt{(x - x')^2 + (y - y')^2}) - K_0 (k \sqrt{(x + x')^2 + (y - y')^2})\right\}.\nonumber \\
\end{eqnarray}
We numerically calculate the integral in Eq. (\ref{delta_t_cal}) as
\begin{eqnarray}
\delta T(x_i,y_j) \simeq \sum_{k,l} \frac{\Delta x \Delta y}{d^2}G(x_k,y_l|x_i,y_j) \tilde{T}(x_k,y_l) - \sum_{l}\Delta y \delta T_{\rm C}(y_l) \frac{\partial G}{\partial x} (x_k,y_l| x_i,y_j)
\end{eqnarray}
with $\Delta x \equiv 5d = |x_i - x_{i+1}|$, $\Delta y = |y_i - y_{i+1}|$ and compared with numerical data for $\delta T(x,y) = T_{\rm g}(x,y) - T_0$ in Fig. \ref{green_func_comp}. 
Here, the value of $T_0$ is the average of $T_{\rm g}(x = 3\Delta x,y)$ along $y$. The inhomogeneity of $\delta T$ appeared in the dead zone rapidly decreases as $x \to \infty$ both for simulation data (i) and Eq. (\ref{delta_t_cal}) (ii). We note that the characteristic length for the inhomogeneity is estimated as $1/kd \simeq 0.696 < \Delta x$. Interestingly, the analytic result in terms of Green's function well agrees with that of our simulation in the inhomogeneous region for $x < 1/k$. This result means that the heat flux from the boundary would not play an important role and the granular temperature is determined through the local shear rate. Furthermore, as discussed in Sec. 3.3, the numerical data for $md^2 D_{xy} / T_{\rm g}$ near $(x, y) \sim (0, 0)$ deviates from Eq. (\ref{tg_dxy}), due to the singularity of the source line. Thus, although there exist the inhomogeneity of the temperature, $T_{\rm g}$ is locally determined through shear rate, i.e. $T_{\rm g} \sim md^2D_{xy}^2$ is still valid, because the gradient of $\sqrt{2T_{\rm g}/m}$ is much smaller than that of the velocity field in our setup.


\end{document}